\documentclass[aps,preprint,epsfig,rotate]{revtex4}
\usepackage{graphicx}
\usepackage{bm}
\usepackage{epsfig}

\begin{document}
\title{Reduction of the canonical Hamiltonian of the metric GR to its natural form}
 \author{Alexei M. Frolov}
 \email[E--mail address: ]{alex1975frol@gmailcom}

\affiliation{Department of Applied Mathematics, \\
 University of Western Ontario, London, Ontario N6H 5B7, Canada}

\date{\today}

\begin{abstract}

The canonical Hamiltonian $H_C$ of the metric General Relativity is reduced to its natural form. The natural form of canonical 
Hamiltonian provides numerous advantages in actual applications to the metric GR, since the general theory of dynamical systems 
with such Hamiltonians is well developed. Furthermore, many analytical and numerically exact solutions for dynamical systems 
with natural Hamiltonians have been found and described in detail. In particular, based on this theory we can discuss an obvious 
analogy between gravitational field(s) and few-particle systems where particles are connected to each other by the Coulomb, or 
harmonic potentials. We also developed an effective method which is used to determine various Poisson brackets between analytical 
functions of the dynamical variables. Furthermore, such variables can be chosen either from the straight, or dual sets of 
symplectic dynamical variables which always arise in any Hamiltonian formulation developed for the metric gravity.  
 
\noindent 
PACS number(s): 04.20.Fy and 11.10.Ef
\end{abstract}

\maketitle
\newpage
\hspace{8.75cm} \textit{Papasha Dirac shevstvuet nad nami} 

\section{Introduction}

In 1958 Dirac published his famous Hamiltonian formulation of the metric General Relativity (or metric gravity, for short) \cite{Dir58}. 
Since then and for a very long time that Dirac's formulation was known as the only correct Hamiltonian approach ever developed for the 
metric gravity. In particular, only by using this Hamiltonian formulation, i.e., the primary and secondary constraints derived in this
Dirac's approach, one was able to restore the complete and correct gauge symmetry (diffeomorphism) of the free (metric) gravitational field(s).
A different Hamiltonian formulation of the metric GR published earlier in \cite{PirSS} was overloaded with numerous mistakes, which can 
easily be found, e.g., in all secondary constraints derived in \cite{PirSS}. Moreover, some important steps of the Hamiltonian procedure, 
developed earlier by Dirac in \cite{Dir50}, were missing in \cite{PirSS}. For instance, the closure of Dirac procedure \cite{Dir50} was 
not demonstrated et al. In reality, it is impossible to show such a closure with wrong secondary constraints, but after reading \cite{PirSS} 
one can get an impression that authors did not understand why they need to do this, in principle. The complete and correct version of the 
Hamiltonian formulation for the metric gravity, originally proposed in \cite{PirSS}, was re-developed and corrected only in 2008 \cite{K&K} 
by Kiriushcheva and Kuzmin. Below, to respect this fact we shall call the Hamiltonian formulaion of the metric GR developed in \cite{K&K} by 
the K$\&$K approach. This approach also allows one to restore the complete diffeomorphism as a correct gauge symmetry of the free gravitational
field. 

Note that after publication \cite{K&K} there were two different and non-contradictory Hamiltonian formulations of the metric gravity. 
Therefore, it was very interesting to investigate relations between these two approaches. In \cite{FK&K} we have shown that Dirac 
formulation of the metric GR and `alternative' K$\&$K-formulation are related to each other by a canonical transformation of dynamical 
variables of the problem, i.e., by a transformation of the generalized `coordinates' $g_{\alpha\beta}$ and corresponding `momenta' 
$\pi^{\mu\nu}$. Furthermore, such a canonical transformation has special and relatively simple form (more details can be found in 
\cite{FK&K}). After an obvious success of our analysis in \cite{FK&K} the following question has imediately arose: is it possible to 
derive another canonical transformation of dynamical variables in the metric gravity which can reduce the canonical Hamiltonian $H_C$ 
of the metric GR derived in \cite{K&K} to some relatively simple forms which are well known in classical mechanics? If the answer is 
`Yes', then it opens access to a large number of analytical and numerical methods developed for classical dynamical systems with such 
Hamiltinians. Furthermore, for many similar systems the corresponding solutions and their properties are also known and we can use 
these solutions to solve `new' gravitational problems, etc. Below, to answer this question we present the new canonical transformation 
of dynamical variables, i.e., generalized coordinates and momenta, in the metric General Relativity. This new canonical transformation 
is also a very special and unique, since it reduces the canonical Hamiltonian $H_C$ of metric GR to the natural form which is almost 
identical to the natural form of many `regular' Hamiltonians already known in analytical mechanics of the potential (dynamical) systems. 
For instance, similar Hamiltonians describe the non-relativistic system of interacting $N$ point particles, where all inter-particle 
forces are generated by some regular potential(s). 

This paper has the following structure. In the next two Sections we introduce the $\Gamma - \Gamma$ Lagrangian ${\cal L}$ of the 
metric General Relativity. By using this Lagrangian ${\cal L}$ we define the corresponding momenta $\pi^{\alpha\beta}$. At the next 
stage of our method we apply the Legendre transformation to exclude velocities and construct the canonical $H_C$ and total $H_t$ 
Hamiltonians of the metric General Relativity. All derived formulas, equations and even logic used in next two Sections are pretty 
standard for any Hamiltonian formulation of the metric GR. Moreover, they were derived and discussed in a number of earlier studies 
(see, e.g., \cite{K&K} and \cite{Fro1}). Nevertheless, the two following Sections are important to make and keep this study 
completely independent of other publications and united by a central idea to illustrate the power of canonical transformations for 
Hamiltonian systems. The fundamental and secondary Poisson brackets are defined and calculated in Section IV. These brackets are the 
main working tools to perform research and obtain solutions for any Hamilton dynamical system, including our Hamiltonian system of 
the gravitational field(s) defined in the metric General Relativity. In particular, our Poisson brackets are used to investigate a 
few fundamental problems currently known in metric GR. Section VI is the central part of this study, since here the canonical 
Hamiltonian $H_C$ of the metric GR is reduced to its natural form. Here we also illustrate a number of advantages of the normal form 
of the canonical Hamiltonian $H_C$ for numerous problems known in the metric GR. A few directions for future development of metric GR 
are also discussed there. Concluding remarks can be found in the last Section. 

Now, let us introduce a few principal notations which are extensively used below. Everywhere in this study we assume that our readers 
are familiar with the tensor calculus, tensor notations and tensor analysis at least at the level of excellent Kochin's book 
\cite{Kochin}. Notations from that book, the rules of tensor transformatons, etc, are used below without any additional reference. In 
particular, in this study the notation $g_{\alpha\beta}$ stands for the covariant components of the metric tensor which are dimensionless 
quantities. Note that all components of the metric tensor $g_{\alpha\beta}$ can be considered either as the actual gravitational fields, 
or as the tensor components of one (united) gravitational field. Each of the $g_{\alpha\beta}$ components is a function of spatial and 
temporal coordinates, i.e., $x^{\alpha} = ( x^{0}, x^{1}, \ldots, x^{d-1})$ in our current notations. In this study all components of 
metric tensor $g_{\alpha\beta}$ are considered as the generalized coordinates of the problem. Analogous notations $\pi^{\alpha\beta}$ 
designate the corresponding contravariant components of momenta which are conjugate to the covariant components $g_{\alpha\beta}$ of the 
metric tensor (see below and references \cite{K&K} and \cite{FK&K}). 

The determinant of the metric tensor $g_{\alpha\beta}$ is denoted by its traditional notation $- g$, where $- g > 0$. The Latin alphabet 
is used for spatial components of vectors/tensors, while the index 0 means their temporal component. In this study the notation $d$ 
(where $d \ge 3$ \cite{X}) designates the total dimension of our space-time manifold. This means that an arbitrary Greek index $\alpha$ 
varies between 0 and $d - 1$, while an arbitrary Latin index varies between 1 and $d - 1$. The quantities and tensors such as 
$B^{((\alpha \beta) \gamma | \mu \nu \lambda)}, I_{mnpq}$, etc, applied below, have been defined in earlier papers \cite{Dir58}, 
\cite{K&K}, \cite{FK&K} and \cite{Fro1}. In this study the definitions of all these quantities and tensors are exactly the same as 
in \cite{K&K} and \cite{FK&K} and there is no need to repeat them. The short notations $g_{\alpha\beta,k}$ and $g_{\gamma\rho,0}$
are used below for the spatial and temporal derivatives, respectively, of the corresponding components of the metric tensor. Any 
expression which contains a pair of identical (or repeated) indexes, where one index is covariant and another is contravariant, 
means summation over this `dummy' index. This convention is very convenient and drastically simplifies many formulas derived in  
metric GR. 

\section{$\Gamma - \Gamma$ Lagrangian of the metric General Relativity}

In this Section we introduce the Lagrangian of the metric General Relativity. Formally, such a Lagrangian (or Lagrangian density) 
should coincide with the integrand in the Einstein-Hilbert integral-action (see, e.g., \cite{LLTF} and \cite{Carm}). However, that 
Lagrangian, which is often called the Einstein-Hilbert Lagrangian, contains a number of derivatives of the second order and cannot 
be used directly in the principle of least action. By applying some standard procedure (see, e.g., \cite{LLTF}) one can transform 
the `singular' Einstein-Hilbert Lagrangian into the `regular' $\Gamma - \Gamma$ Lagrangian which contains no second order derivative 
and is written in the form
\begin{eqnarray}
  {\cal L} &=& \frac14 \sqrt{-g} B^{\alpha\beta\gamma\mu\nu\rho} \Bigl(\frac{\partial g_{\alpha\beta}}{\partial x^{\gamma}}\Bigr) 
 \Bigl(\frac{\partial g_{\mu\nu}}{\partial x^{\rho}}\Bigr) = \frac14 \sqrt{-g} B^{\alpha\beta\gamma\mu\nu\rho} g_{\alpha\beta,\gamma} 
 g_{\mu\nu,\rho} \label{eq05} 
\end{eqnarray}
where 
\begin{eqnarray}
 B^{\alpha\beta\gamma\mu\nu\rho} &=& g^{\alpha\beta} g^{\gamma\rho} g^{\mu\nu} - g^{\alpha\mu} g^{\beta\nu} g^{\gamma\rho} + 2 
g^{\alpha\rho} g^{\beta\nu} g^{\gamma\mu} - 2 g^{\alpha\beta} g^{\gamma\mu} g^{\nu\rho} \; \; \label{Bcoef}
\end{eqnarray}
is a homogeneous cubic function of the contravariant components of the metric tensor $g^{\alpha\beta}$. This formula can also be written
as the cubic function of the inverse powers of covariant components of the metric tensor $g_{\alpha\beta}$. The both forms of the 
$B^{\alpha\beta\gamma\mu\nu\rho}$ tensor are equivalent, since the equality $g_{\alpha\gamma} g^{\gamma\beta} = g^{\alpha}_{\beta} = 
\delta^{\alpha}_{\beta}$ is always obeyed \cite{Kochin}. In this study the covariant components of the metric tensor $g_{\alpha\beta}$ 
are chosen as the straight set of coordinates for the Hamiltonian formulation(s) of the metric GR. In thjis case, the contravariant 
components of the metric tensor $g^{\alpha\beta}$  form the corresponding set of dual coordinates. For tensor Hamiltonian fields these 
two sets of coordinates (in fact, the two sets of canonical variables which include these coordinates) are very closely related to 
each other by the Poisson brackets (see discussion below). Note also that in the right-hand side of this formula, Eq.(\ref{eq05}), the 
short notation $g_{\alpha\beta,\gamma}$ designates the partial derivatives $\frac{\partial g_{\alpha\beta}}{\partial x^{\gamma}}$ in 
respect to the spatial/temporal coordinates. Note that the $\Gamma - \Gamma$ Lagrangian ${\cal L}$, Eq.(\ref{eq05}), contains the partial 
temporal derivatives $g_{0 \sigma,0} (= g_{\sigma 0,0})$ of the first-order only, and it is used below to derive the total Hamiltonian of 
the metric GR. In some papers the temporal derivatives $g_{0 \sigma,0}$ were called the $\sigma$-velocities. 

In reality, to derive the closed formula for the Hamiltonian of metric GR we need a slightly different form of the $\Gamma - \Gamma$ 
Lagrangian where all temporal derivatives (or time-derivatives) are explicitly separated from other derivatives (see, 
e.g., \cite{K&K})  
\begin{eqnarray}
  {\cal L} = \frac14 \sqrt{-g} B^{\alpha\beta 0\mu\nu 0} g_{\alpha\beta,0} g_{\mu\nu,0} + \frac12 \sqrt{-g} B^{(\alpha\beta 0|\mu\nu k)} 
  g_{\alpha\beta,0} g_{\mu\nu,k} + \frac14 \sqrt{-g} B^{\alpha\beta k \mu\nu l} g_{\alpha\beta,k} g_{\mu\nu,l} \label{eq51}
\end{eqnarray}
where the notation $B^{(\alpha\beta\gamma|\mu\nu\rho)}$ means a `symmetrical' $B^{\alpha\beta\gamma\mu\nu\rho}$ quantity which is 
symmetrized in respect to the permutation of two groups of indexes, i.e.,
\begin{eqnarray}
 B^{(\alpha\beta\gamma|\mu\nu\rho)} &=& \frac12 \Bigl( B^{\alpha\beta\gamma\mu\nu\rho} + B^{\mu\nu\rho\alpha\beta\gamma} \Bigr) 
 = g^{\alpha\beta} g^{\gamma\rho} g^{\mu\nu} - g^{\alpha\mu} g^{\beta\nu} g^{\gamma\rho} \nonumber \\ 
 &+& 2 g^{\alpha\rho} g^{\beta\nu} g^{\gamma\mu} - g^{\alpha\beta} g^{\nu\rho} g^{\gamma\mu} - g^{\alpha\rho} 
 g^{\beta\gamma} g^{\mu\nu}  \;  \label{eq52}
\end{eqnarray}

By using the Lagrangian ${\cal L}$, Eq.(\ref{eq51}), and standard definition of momentum as a partial derivative of the Lagrangian in
respect to the corresponding velocity (see, e.g., \cite{Dir64}), we obtain the explicit formulas for all components of the tensor of 
momentum $\pi^{\gamma\sigma}$   
\begin{eqnarray}
  \pi^{\gamma\sigma} = \frac{\partial {\cal L}}{\partial g_{\gamma\sigma,0}} = \frac{1}{2} \sqrt{-g} B^{((\gamma\sigma) 0|\mu\nu 0)} 
  g_{\mu\nu, 0} + \frac{1}{2} \sqrt{-g} B^{((\gamma\sigma) 0|\mu\nu k)} g_{\mu\nu, k} \; \; \; \label{mom}
\end{eqnarray}
The first term in the right-hand side of the last equation can be written in the form 
\begin{eqnarray}
 \frac{1}{2} \sqrt{-g} B^{((\gamma\sigma)0|\mu\nu 0)} g_{\mu\nu, 0} = \frac{1}{2} \sqrt{-g} g^{00} E^{\mu\nu\gamma\sigma} g_{\mu\nu, 0} 
 \; \; \label{B}
\end{eqnarray}
where the Dirac tensors $E^{\mu\nu\gamma\sigma}$ and $e^{\mu \nu}$ are
\begin{eqnarray}
 E^{\mu \nu \gamma \rho} = e^{\mu \nu} e^{\gamma \rho} - e^{\mu \gamma} e^{\nu \rho} \; \; , \; \; {\rm and} \; \; \; e^{\mu \nu} = 
 g^{\mu \nu} - \frac{g^{0 \mu} g^{0 \nu}}{g^{00}} \; \; \; \label{E}  
\end{eqnarray}
and it is easy to check that $E^{\mu\nu\gamma\sigma} = E^{\gamma\sigma\mu\nu}$ and $e^{\mu \nu} = e^{\nu \mu}$. Also, as follows 
directly from the formula, Eq.(\ref{E}), the tensor $e^{\mu \nu}$ equals zero, if either index $\mu$, or index $\nu$ (or both) equals 
zero. The same statement is true for the Dirac $E^{\mu\nu\gamma\sigma}$ tensor, i.e., $E^{0\nu\gamma\sigma} = 0, E^{\mu 0\gamma\sigma} 
= 0, E^{\mu\nu 0\sigma} = 0$ and $E^{\mu\nu\gamma 0} = 0$. The $E^{pqkl}$ quantity is called the space-like Dirac tensor of the fourth 
rank. Note that all components of this space-like tensor $E^{p q k l}$ are not equal zero. Furthermore, the space-like tensor 
$E^{p q k l}$ is a positively-defined and invertable tensor. Its inverse space-like tensor $I_{m n p q}$ is also positively-defined 
and invertable space-like tensor of the fourth rank which is written in the form
\begin{equation}
 I_{m n q p} = \frac{1}{d - 2} g_{m n} g_{p q} -  g_{m p} g_{n q} \label{I}
\end{equation}
This tensor plays a very important role in our Hamiltonian analysis (see below). From here we can write $I_{m n p q} E^{p q k l} = 
g^{k}_{m} g^{l}_{n} = \delta^{k}_{m} \delta^{l}_{n}$, where the $g^{\alpha}_{\beta} = \delta^{\alpha}_{\beta}$ tensor is the substitution 
tensor \cite{Kochin}, while the symbol $\delta^{\alpha}_{\beta}$ denotes the Kroneker delta (it equals zero for all possible indexes, 
unless $\alpha = \beta$, when its numerical value equals unity).

First, let us consider the `regular' case when in Eq.(\ref{mom}) $\gamma = p$ and $\sigma = q$. In this case one finds the following 
formulas for double space-like components of the momentum tensor
\begin{eqnarray}
  \pi^{pq} = \frac{\partial {\cal L}}{\partial g_{p q,0}} = \frac{1}{2} \sqrt{-g} B^{((p q) 0|\mu\nu 0)} g_{\mu\nu,0} + \frac{1}{2} 
 \sqrt{-g} B^{((p q) 0|\mu\nu k)} g_{\mu\nu, k} \; \; \label{momenta}
\end{eqnarray}
For each pair of $(pq)-$indexes (or $(mn)-$indexes). The tensor in the right-hand side of this equation is invertable and the velocity 
$g_{m n, 0}$ is explicitly expressed as the linear function (or linear combination) of the space-like components $\pi^{pq}$ of momentum 
tensor:
\begin{eqnarray}
  g_{mn, 0} &=& \frac{1}{g^{00}} \Bigl( \frac{2}{\sqrt{-g}} I_{m n p q} \pi^{pq} - I_{m n p q} B^{((pq) 0|\mu\nu k)} g_{\mu\nu, k} \Bigr)
 \nonumber \\ 
 &=& \frac{1}{g^{00}} I_{m n p q} \Bigl( \frac{2}{\sqrt{-g}} \pi^{pq} - B^{((pq) 0|\mu\nu k)} g_{\mu\nu, k} \Bigr) \label{veloc}
\end{eqnarray}
where the Dirac tensor $I_{m n p q}$ is defined by Eq.(\ref{I}). As follows from Eqs.(\ref{momenta}) and (\ref{veloc}) for the space-like 
components of metric tensor $g_{pq}$ and corresponding momenta $\pi^{mn}$ one finds no principal difference with the Hamilton dynamical 
systems, which are routinely studied in classical mechanics. Indeed, these space-like components of momenta and corresponding velocities 
are related to each other by a very simple (linear) equation. However, even these components of momenta $\pi^{pq}$ do not related with 
the corresponding velocities $g_{pq,0}$ directly, i.e., by one equation and/or by one scalar parameter, e.g., by some `effective' mass. 
Instead, for gravitational field(s) the corresponding relation, Eq.(\ref{veloc}), has a matrix form and one space-like component of 
momenta $\pi^{mn}$ depends upon quasi-linear combination \cite{QL} of different velocities $g_{pq,0}$ (and vice versa). Nevertheless, even 
such a `non-traditional' matrix definition of momenta works very well in actual applications and, in particular, allows one to develop the 
complete and non-cotradictive Hamiltonian approach for the metric GR.

In the second `non-regular' (or singular) case, when $\gamma = 0$, the first term in the right-hand side of Eq.(\ref{mom}) equals zero 
and this equation takes the from 
\begin{eqnarray}
  \pi^{0\sigma} = \frac{\partial {\cal L}}{\partial g_{0\sigma,0}} = \frac{1}{2} \sqrt{-g} B^{((0\sigma) 0|\mu\nu k)} g_{\mu\nu, k} 
   \; \; \; \label{constr}
\end{eqnarray}
which contains no velocity et al. Furthermore, this equation, Eq.(\ref{constr}), determines the momentum $\pi^{0\sigma}$ as a polynomial 
(cubic) functions of the contravariant components of the metric tensor $g^{\alpha\beta}$ and a linear function of the both $\sqrt{- g}$ 
value and spatial derivatives of the covariant components $g_{\mu\nu, k}$ of metric tensor. It is clear that such a situation cannot be 
found neither in classical mechanics, nor in quantum mechanics of arbitrary systems of particles. However, for actual physical fields 
similar situations arise quite often. The physical meaning of Eq.(\ref{constr}) is simple and can be expressed in the following words. 
The function
\begin{eqnarray}
   \phi^{0\sigma} = \pi^{0\sigma} - \frac{1}{2} \sqrt{-g} B^{((0\sigma) 0|\mu\nu k)} g_{\mu\nu, k} \; \label{primary}
\end{eqnarray}
must be equal zero at any time, i.e., it does not change during actual physical motions (or time-evolution) of the gravitational field. 
Dirac in \cite{Dir50} proposed to write such equalities in the symbolic form $\phi^{0\sigma} \approx 0$ and called these $d$ functions 
$\phi^{0\sigma}$ (for $\sigma = 0, 1, \ldots, d - 1$), Eq.(\ref{primary}), by the primary constraints (see also \cite{Dir64}).

\section{Total and canonical Hamiltonians of metric General Relativity}

Now, by applying the Legendre transformation to the known $\Gamma - \Gamma$ Lagrangian ${\cal L}$, of the metric GR, Eq.(\ref{eq51}), 
and excluding all space-like field-velocities $g_{mn,0}$ we can derive the following formulas for the total and canonical Hamiltonians 
of the metric GR. In particular, the total Hamiltonian $H_t$ of the gravitational field in metric GR derived from the $\Gamma - \Gamma$ 
Lagrangian ${\cal L}$, Eq.(\ref{eq05}), is written in the form 
\begin{eqnarray}
  H_t = \pi^{\alpha\beta} g_{\alpha\beta,0} - {\cal L} = H_C + g_{0\sigma,0} \phi^{0\sigma}  \label{eq1}
\end{eqnarray}
where $\phi^{0\sigma} = \pi^{0\sigma} - \frac{1}{2}\sqrt{-g} B^{\left( \left(0\sigma\right) 0\mid\mu\nu k\right)} g_{\mu\nu,k}$ are 
the primary constraints, while $g_{0\sigma,0}$ are the corresponding $\sigma-$velocities' and $H_C$ is the canonical Hamiltonian of 
metric GR
\begin{eqnarray}
 & &H_C = \frac{1}{\sqrt{-g} g^{00}} I_{mnpq} \pi^{mn} \pi^{pq} - \frac{1}{g^{00}} I_{mnpq} \pi^{mn} B^{(p q 0|\mu \nu k)} 
 g_{\mu\nu,k} \label{eq5} \\
 &+& \frac14 \sqrt{-g} \Bigl[ \frac{1}{g^{00}} I_{mnpq} B^{((mn)0|\mu\nu k)} B^{(pq0|\alpha\beta l)} - 
 B^{\mu\nu k \alpha\beta l}\Bigr] g_{\mu\nu,k} g_{\alpha\beta,l} \nonumber
\end{eqnarray}
which does not contain any primary constraint $\phi^{0\sigma}$. All $d$ primary constraints $\phi^{0\sigma}$, where $\sigma = 0, 1, \ldots, 
d - 1$, are included in the total Hamiltonian $H_t$, Eq.(\ref{eq1}). It should be emphasized again that these primary constraints arise during 
our transition from the $\Gamma - \Gamma$ Lagrangian ${\cal L}$, Eq.(\ref{eq05}), to the Hamiltonians $H_t$ and $H_C$, since the $\Gamma - 
\Gamma$ Lagrangian ${\cal L}$ is a linear (not quadratic!) function of all $d$ momenta $\pi^{0\sigma} = \frac{\delta L}{\delta 
g_{0\sigma,0}}$ each of which includes at least one temporal index \cite{K&K}. The total and canonical Hamiltonians $H_t$ and $H_C$ are the 
scalar functions defined in the $2 d-$dimensional phase space $\Bigl\{ g_{\alpha\beta}, \pi^{\mu\nu} \Bigr\}$, where components of the 
metric $g_{\alpha\beta}$ tensor and momentum tensor $\pi^{\mu\nu}$ have been chosen as the basic dynamical variables. Such a phase space 
is, in fact, a symplectic space and the corresponding symplectic structure is determined by the Poisson brackets between its basic dynamical 
variables, i.e., coordinates $g_{\alpha\beta}$ and momenta $\pi^{\mu\nu}$. Now we need to define the Poisson brackets (or commutators) which 
play a great role in any the Hamiltonian formulation developed for the metric GR. These Poisson brackets are introduced in the next Section.   

\section{Poisson brackets}

Let us define the Poisson brackets (or PB, for short) which are absolutely crucial for the creation, development and applications of any 
Hamiltonian-based approach in the metric General Relativity. From now on we shall consider only Hamiltonian approaches (in metric GR) which 
are canonically related either to the K$\&$K-approach \cite{K&K}, or to the Dirac approach \cite{Dir58}. Note again that these two 
Hamiltonian formulations are canonically related to each other (for more details, see \cite{FK&K}). Therefore, it is possible to obtain and 
present the basic (or fundamental) set of Poisson brackets only for one of these two Hamiltonian formulations, e.g., for the K$\&$K-approach. 
Analogous Poisson brackets for other Hamiltonian formulations of metric GR can be derived from these `fundamental' values known in the 
K$\&$K-approach. The basic Poisson brackets between $\frac{d(d + 1)}{2}$ components of the momentum tensor $\pi^{\mu\nu}$ and $\frac{d(d + 
1)}{2}$ `coordinates' $g_{\alpha\beta}$ in the K$\&$K-approach are \cite{K&K}
\begin{eqnarray}
  [ g_{\alpha\beta}, \pi^{\mu\nu}] = - [ \pi^{\mu\nu}, g_{\alpha\beta}] = g_{\alpha\beta} \pi^{\mu\nu} - \pi^{\mu\nu} g_{\alpha\beta} 
  = \frac12 \Bigl(g^{\mu}_{\alpha} g^{\nu}_{\beta} + g^{\nu}_{\alpha} g^{\mu}_{\beta}\Bigr) = \frac12 \Bigl(\delta^{\mu}_{\alpha} 
  \delta^{\nu}_{\beta} + \delta^{\nu}_{\alpha} \delta^{\mu}_{\beta}\Bigr) = \Delta^{\mu\nu}_{\alpha\beta} \; \; \; ,  \label{eq15} 
\end{eqnarray}
where $g^{\mu}_{\alpha} = \delta^{\mu}_{\alpha}$ is the substitution tensor \cite{Kochin} and symbol $\delta^{\mu}_{\beta}$ is the Kronecker 
delta, while the notation $\Delta^{\mu\nu}_{\alpha\beta}$ stands for the gravitational (or tensor) delta-function. All other fundamental 
Poisson brackets between basic dynamical variables of the metric GR equal zero identically, i.e., $[ g_{\alpha\beta}, g_{\mu\nu}] = 0$ and 
$[ \pi^{\alpha\beta}, \pi^{\mu\nu}] = 0$. This set of $\frac{d^{2}(d^{2} - 1)}{4}$ Poisson brackets has a fundamental value, since these 
PB define the unique symplectic structure directly related to the Rimanian structure of the original $d (d + 1)$-dimensional tensor phase 
space and to the metric tensor $g_{\alpha\beta}$. We hope that readers are familiar with the general properties of Poisson brackets (see, 
e.g., \cite{Gant} - \cite{GF}).  

In general, the $\frac{d^{2}(d^{2} - 1)}{4}$ Poisson brackets mentioned above are sufficient to operate successfully in any correct Hamiltonian 
approach developed for the metric GR. However, in many applications it is crucially important to determine other Poisson brackets, which are also 
called the secondary PB. The secondary PB are calculated between different analytical functions of basic dynamical variables, i.e., coordinates 
and momenta, but they arise quite often in actual calculations. In general, it is difficult and time-consuming to derive the explicit formulas for 
secondary PB every time when you need them. Furthermore, in actual applications one usually needs to determine a few hundreds of different Poisson 
brackets. Here we present a number of additional (or secondary) Poisson brackets which are sufficient for our purposes in this study. The first 
additional group of secondary Poisson brackets is   
\begin{eqnarray}
  [ g^{\alpha\beta}, \pi^{\mu\nu}] = - \frac12 \Bigl( g^{\alpha\mu} g^{\beta\nu} + g^{\alpha\nu} g^{\beta\mu} \Bigr) \; \; {\rm and} \; 
  \; [ g^{\alpha\beta}, g_{\mu\nu}] = 0 \; . \label{eq151} 
\end{eqnarray}
which include the contravariant components of the metric tensor $g^{\alpha\beta}$. Note that the $g^{\alpha\beta}$ tensor is inverse of the
$g_{\alpha\beta}$ tensor, since the following equations $g_{\alpha\gamma} g^{\gamma\beta} = g_{\alpha}^{\beta} = \delta_{\alpha}^{\beta} = 
g^{\beta\gamma} g_{\gamma\alpha}$ are always obeyed between components of the metric tensor. Therefore, we need to check the correctness of 
Eq.(\ref{eq151}) in the case of direct replacement $g^{\alpha\beta} \rightarrow \frac{1}{g^{\alpha\beta}}$. The second sub-equation in 
Eq.(\ref{eq151}), i.e., $[ \frac{1}{g^{\alpha\beta}}, g_{\mu\nu}] = 0$ does not change its form, while for the first sub-equation one finds
\begin{eqnarray}
 [ g^{\alpha\beta}, \pi^{\mu\nu}] = [ \frac{1}{g_{\alpha\beta}}, \pi^{\mu\nu}] = - \Bigl(\frac{1}{g_{\alpha\beta}}\Bigr)^{2} 
 [ g_{\alpha\beta}, \pi^{\mu\nu}] = - (g^{\alpha\beta})^{2} \Delta^{\mu\nu}_{\alpha\beta} = - \frac12 \Bigl( g^{\alpha\mu} g^{\beta\nu} + 
 g^{\alpha\nu} g^{\beta\mu} \Bigr) \; \; , \label{eq151a} 
\end{eqnarray}
which coincides with the first equality in Eq.(\ref{eq151}) and we do not have any contradiction here. 
  
The second set of additional Poisson brackets arises, if one explicitly introduces the dual system of dynamical variables $\{ g^{\alpha\beta}, 
\pi_{\mu\nu}\}$ which always exists for any tensor Hamiltonian system. When I started to write this paper one of my goals was to avoid the 
use of components of the `dual momentum' $\pi_{\mu\nu}$ as dynamical variables. However, after a number of attempts I gave up and arrived to 
the following conclusion: to create a truly correct and non-contradictory Hamiltonian formulation for some dynamical tensor system we have to 
deal with the two different $d (d + 1)-$dimensional sets of dynamical variables: (a) the straight set $\{ g_{\alpha\beta}, \pi^{\mu\nu}\}$, 
and (b) the dual set $\{ g^{\alpha\beta}, \pi_{\mu\nu}\}$. The Poisson brackets between all dynamical variables from these two sets must be 
derived and carefully checked for non-contradictory. In those cases when all these Poisson brackets (for dynamical variables from the straight 
and dual sets) do not contradict each other we can say that our newly created Hamiltonian formulation is truly covariant, self-sustained and 
correct. Otherwise, one needs to re-define all momenta and try to repeat the whole Hamilton procedure from the very beginning. The necessity 
to deal with the two sets of dynamical variables instantaneously is an important difference between Hamiltonian procedures developed for the 
affine vector spaces and Riemanian tensor spaces. In other words, the instant presence of two sets of dynamical variables (straight and dual 
sets) is a common feature of all Hamiltonian formulations for the tensor fields. It can be shown that only by dealing with the both straight 
and dual sets of dynamical variables we can guarantee the internal covariance and self-sustainability of our Hamiltonian approach developed 
for the metric GR. 

The fact that we need to operate with the both straight and dual systems of dynamical variables in any Hamiltonian formulation developed for 
tensor dynamical systems can be illustrated by the following example. Let us suppose that we have defined the momentum as above, i.e., we 
introduced the contravariant tensor of momentum $\pi^{\rho\sigma}$. Then, by using the metric tensor $g_{\alpha\beta}$ we can introduce the 
new tensor of momentum $\pi_{\mu\nu} = g_{\mu\rho} g_{\nu\sigma} \pi^{\rho\sigma} = g_{\mu\rho} \pi^{\rho\sigma} g_{\nu\sigma} = 
\pi^{\rho\sigma} g_{\mu\rho} g_{\nu\sigma}$ which is a covariant tensor of second rank. The same transition ($\pi^{\rho\sigma} \rightarrow 
\pi_{\rho\sigma}$) changes the corresponding Poisson brackets. Some terms in the `new' PB are transformed easily, while analogous transformations 
for other terms are hard to find. Nevertheless, all these new PB must be determined correctly. Arguments such as `we do not want to introduce this 
new tensor of momentum' cannot be considered as serious, since, if the momentum $\pi^{\rho\sigma}$ is a true contravariant tensor, then it should 
be transformed as a tensor. In reality, someone can take the  covariant components of this new momentum $\pi_{\mu\nu}$ as the new  $\frac{d(d + 
1)}{2}$ dynamical variables. The corresponding coordinates in this `new' Hamiltonian formulation are chosen as components of the contravariant 
$g^{\alpha\beta}$ metric tensor. Briefly, these dynamical variables $\{g^{\alpha\beta}, \pi_{\rho\sigma} \}$ lead to another `new' Hamiltonian 
formulation of the metric GR. It is clear that the both Hamiltonian formulations developed with these two sets of basic dynamical variables must 
essentially be the same, or at least, they must be related to each other by a canonical transformation (otherwise, both of them are wrong). Let us 
present the Poisson brackets for the dual set of dynamical variables $\{ g^{\alpha\beta}, \pi_{\mu\nu}\}$
\begin{eqnarray}
  [ g_{\alpha\beta}, \pi_{\mu\nu}] = \frac12 \Bigl( g_{\alpha\mu} g_{\beta\nu} + g_{\alpha\nu} g_{\beta\mu} \Bigr) \; \; {\rm and} \; \;
  [ g^{\alpha\beta}, \pi_{\mu\nu}] = - \frac12 \Bigl( g^{\alpha}_{\mu} g^{\beta}_{\nu} + g^{\alpha}_{\nu} g^{\beta}_{\mu} \Bigr) = 
  - \Delta^{\alpha\beta}_{\mu\nu} \; \; . \label{eq153} 
\end{eqnarray}
and also $[ g^{\alpha\beta}, g^{\mu\nu}] = 0 , [ \pi_{\alpha\beta}, \pi_{\mu\nu}] = 0$ and $[ g_{\alpha\beta}, g^{\mu\nu}] = 0$. The last PB
bracket which we want to present here is 
\begin{equation}
  [ \pi_{\alpha\beta}, \pi^{\mu\nu}] = \pi_{\alpha}^{\mu} \delta_{\beta}^{\nu} + \delta_{\alpha}^{\mu} \pi_{\beta}^{\nu} \; \; \; , \; \; 
  \label{pipi}
\end{equation}
This means that the co- and contra-covariant components of the momentum tensor do not commute with each other. By using these Poisson brackets 
one can show that the both straight and dual sets of dynamical variables produce almost identical Hamiltonian formulations of metric gravity. 
This means that each of these two Hamiltonian formulation of the metric GR (in the straight and dual spaces) is correct.   

Now, let us present a few following Poisson brackets which are very useful in actual calculations. Let $g (> 0)$ will be the determinant of the 
metric tensor $g_{\alpha\beta}$ and $F(g)$ is an arbitrary analytical function of $g$. In this notation one finds
\begin{eqnarray}
  [ F(g), \pi^{\alpha\beta}] = \Bigl( \frac{\partial F}{\partial g} \Bigr) g g^{\alpha\beta} \; \; \; {\rm and} \; \; \; 
  [ \sqrt{- g}, \pi^{\alpha\beta}] = - \frac{1}{2 \sqrt{- g}} g g^{\alpha\beta} = \frac12 \sqrt{- g} g^{\alpha\beta} \; , \; \label{eq154} 
\end{eqnarray}
for $F(g) = \sqrt{- g}$, if the determinant $g$ is negative. Analogously, for the $\pi_{\alpha\beta}$ momentum we obtain 
\begin{eqnarray}
  [ F(g), \pi_{\alpha\beta}] = \Bigl( \frac{\partial F}{\partial g} \Bigr) g g_{\alpha\beta} \; \; \; {\rm and} \; \; \; 
  [ \sqrt{- g}, \pi_{\alpha\beta}] = - \frac{1}{2 \sqrt{- g}} g g_{\alpha\beta} = \frac12 \sqrt{- g} g_{\alpha\beta} \; \label{eq155} 
\end{eqnarray}
These formulas lead to the following expressions 
\begin{eqnarray}
  [ \frac{1}{\sqrt{- g}}, \pi^{\alpha\beta}] = - \frac{1}{2 \sqrt{- g}} g^{\alpha\beta} \; \; \; {\rm and} \; \; \; 
  [ \frac{1}{\sqrt{- g}}, \pi_{\alpha\beta}] = - \frac{1}{2 \sqrt{- g}} g_{\alpha\beta} \; \label{eq155a} 
\end{eqnarray}
which are important for our calculations performed in the next Sections. All other Poisson brackets needed in calculations can be determined 
with the use of our PB presented in Eqs.(\ref{eq15}) - (\ref{eq155a}). A large number of Poisson brackets which are often needed in various 
problems of metric GR can be found in our paper \cite{FroUnp}.

Another example is slightly more complicated and includes the tensor(s) $e^{\mu \nu}$ defined above. From the explicit formulas for the 
components of $e^{\mu \nu}$ tensor, Eq.(\ref{E}), one finds that only non-zero elements of this tensor are located in the space-like 
corner of the total $e^{\mu \nu}$ tensor. These non-zero elements form the space-like $e^{pq}$ tensor (or space-like part of the total 
$e^{\mu \nu}$ tensor) which is often called the space-like Dirac tensor (or space-like tensor of the second rank). For this tensor one 
easily finds the following useful relation
\begin{eqnarray}
  g_{\alpha\beta} e^{\alpha\beta} = g_{\alpha\beta} g^{\alpha\beta} - g_{\alpha\beta} \Bigl(\frac{g^{\alpha 0} g^{\beta 0}}{g^{00}}\Bigr) 
 = d - g_{\beta}^{0} \; \frac{g^{\beta 0}}{g^{00}} = d - \frac{g^{00}}{g^{00}} = d - 1 = g_{mn} e^{mn} \; \; \label{d-1}
\end{eqnarray}
where $g_{\alpha\beta} g^{\alpha\beta} = d$ and $d$ is the total dimension of our space-time continuum. By using our formulas for the Poisson 
brackets obtained above we derive the following formulas 
\begin{eqnarray}
 [ e^{pq}, \pi^{\alpha\beta}] &=& - \frac12 \Bigl( g^{p\alpha} g^{q\beta} + g^{p\beta} g^{q\alpha} \Bigr) + \frac12 \Bigl( g^{0\alpha} 
 g^{p\beta} + g^{0\beta} g^{p\alpha} \Bigr) \Bigl(\frac{g^{0q}}{g^{00}}\Bigr) \nonumber \\
 &+& \frac12 \Bigl(\frac{g^{0p}}{g^{00}}\Bigr) \Bigl( g^{0\alpha} g^{q\beta} + g^{0\beta} g^{q\alpha} \Bigr) - \frac{g^{0p} g^{q\alpha} 
 g^{0\beta} g^{0q}}{(g^{00})^2} \; \label{e-tens}
\end{eqnarray}  
and 
\begin{eqnarray}
 [ e^{pq}, \pi_{\alpha\beta}] = - \Delta^{pq}_{\alpha\beta} + \Delta^{0 p}_{\alpha\beta} \Bigl(\frac{g^{0q}}{g^{00}}\Bigr)  
 + \frac12 \Bigl(\frac{g^{0p}}{g^{00}}\Bigr) \Delta^{0 q}_{\alpha\beta} - \Delta^{0 0}_{\alpha\beta} \frac{g^{0p} g^{0q}}{(g^{00})^2} 
 \; \label{e-tensA}
\end{eqnarray}  
Analytical formulas for these PB are important, since there were some ideas to use components of this space-like tensor $e^{pq}$ as the new
$\frac{d(d - 1)}{2}$ canonical variables (new coordinates) for another `advanced' Hamiltonian formulation of the metric GR. As follows from 
Eqs.(\ref{e-tens}) and (\ref{e-tensA}) the complexity of arising Poisson brackets makes this idea unworkable. 

To conclude this Section let us present the following formula for the fundamental Poisson brackets written in the united form for the both 
straight and dual stes of dynamical variables  
\begin{eqnarray}
  [ g_{\alpha\beta}, \pi^{\mu\nu}] = \Delta^{\mu\nu}_{\alpha\beta} = [ \pi_{\alpha\beta}, g^{\mu\nu}]  \; \; \; .  \label{eq1551} 
\end{eqnarray}
This beatiful formula includes two fundamental Poisson bracket(s) and clearly shows the differences which arise during transition from the 
straight set of canonical variables to analogous dual set. As follows from the formula, Eq.(\ref{eq155}), the truly dual system of dynamical 
variables (for the original $\{ g_{\alpha\beta}, \pi^{\mu\nu}\}$ system) must be $\{ -g^{\alpha\beta}, \pi_{\mu\nu}\}$ system rather then our 
dual $\{ g^{\alpha\beta}, \pi_{\mu\nu}\}$ system of variables introduced above. Below, we shall ignore this comment and consider the $\{ 
g_{\alpha\beta}, \pi^{\mu\nu}\} \rightarrow \{ g^{\alpha\beta}, \pi_{\mu\nu}\}$ transition as a canonical transformation of dynamical 
variables for our Hamiltonian formulation of the metric GR. Therefore, based on the general theory described in \cite{Gant} we can write the 
following equality 
\begin{eqnarray} 
 \pi^{\mu\nu} \delta g_{\mu\nu} - H_t \delta t + \delta F = v \Bigl( \pi_{\mu\nu} \delta g^{\mu\nu} - \overline{H}_t \delta t \Bigr) 
 \; \; , \; \; \label{eq1553}
\end{eqnarray}
where $v$ is a real, non-zero number which is called the valence of this canonical transformation, while $F(t, g_{\alpha\beta}, 
\pi^{\gamma\sigma})$ is its generating function. The notations $H_t$ and $\overline{H}_t$ means the total Hamiltonians 
written in the both systems of dynamical variables, i.e., in the straight $\{ g_{\alpha\beta}, \pi^{\mu\nu}\}$ and dual $\{ g^{\alpha\beta}, 
\pi_{\mu\nu}\}$ systems of variables, respectively. It is clear that for such a canonical transformation we can use the same time $t$ (for 
both systems) and this transformation is univalent which means that $| v | = 1$ (in reality, we have found that $v = - 1$). Furthermore, it 
is possible to show that for the $\{ g_{\alpha\beta}, \pi^{\mu\nu}\} \rightarrow \{ g^{\alpha\beta}, \pi_{\mu\nu}\}$ canonical transformation 
the generating function $F$ can be chosen in a very special form $F = S(t, g_{\mu\nu}, g^{\alpha\beta})$ which corresponds to the free 
canonical transformation(s). In this case the previous equation takes the form  
\begin{eqnarray} 
 \pi^{\mu\nu} \delta g_{\mu\nu} - H_t \delta t + \delta S(t, g_{\mu\nu}, g^{\alpha\beta}) = v \Bigl( \pi_{\mu\nu} \delta g^{\mu\nu} 
 - \overline{H}_t \delta t \Bigr) \; \; \; \label{eq1555}
\end{eqnarray}
and three following equations are also obeyed (for $v = - 1$)
\begin{eqnarray} 
 \pi^{\mu\nu} = \frac{\partial S}{\partial g_{\mu\nu}} \; \; , \; \; \pi_{\mu\nu} = - \frac{\partial S}{\partial g^{\mu\nu}} \; \; {\rm 
 and} \; \; \overline{H}_t = - H_t + \frac{\partial S}{\partial t} \; \; . \; \; \label{eq1557}
\end{eqnarray}
The last equation, Eq.(\ref{eq1557}), opens a short way to the Jacobi equation for the gravitational field in metric GR, but here we cannot 
discuss this interesting problem (more details can be found in \cite{Fro1}), since it is located outside of the main stream of our current 
analysis. 

\section{Applications of Poisson brackets to actual problems of metric GR}

The knowledge of all Poisson brackets derived above allows one to achieve a number of goals in the Hamiltonian formulation(s) of metric General
Relativity. In particular, by using these Poisson brackets we can complete the actual Hamiltonian formulation of the metric GR. Another problem 
which can be solved with the use of our Poisson brackets is explicit derivation of the Hamilton equations of motion for actual gravitational 
field(s) which are often called the time-evolution equations. Also, with these Poisson brackets we can find the new canonical transformations 
which are simplify either the canonical Hamiltonian $H_C$, or secondary constraints $\chi^{0\sigma}$ (they are defined below). In particular, 
below, we consider the reduction of the canonical Hamiltonian $H_C$ to its natural form. The first two of the mentioned problems are briefly 
considered in the next two subsections. These two problems were extensively discussed in earlier studies \cite{K&K}, \cite{FK&K} and \cite{Fro1}. 
Therefore, there is no need for us here to move into deep analysis of these problems and repeat all formulas derived in those works. Here we just 
want to illustrate how our formulas for Poisson brackets allow one to simplify analytical calculations of many difficult expressions. In contrast 
with this, the third problem (i.e., reduction of $H_C$ to its natural form) is the central part of this study and we have to disclose all details 
of our computations. These details can be found in the next Section. In general, analytical computations of a large number of Poisson brackets is 
a very good exercise in tensor calculus. 
 
\subsection{Constraints and Dirac closure of the Hamiltonian procedure}

Let us complete the Hamiltonian formulation of the metric GR, described above, by using the momenta $\pi^{mn}$, primary constraints $\phi^{0\sigma}$
and canonical Hamiltonian $H_C$ defined in Eq.(\ref{momenta}), Eq.(\ref{constr}) and Eq.(\ref{eq5}), respectively. First, we need to determine 
commutators between the canonical Hamiltonian $H_C$, Eq.(\ref{eq5}), and primary constraints $\phi^{0\sigma}$, Eq.(\ref{primary}). This directly 
leads (see discussion in \cite{K&K}) to the secondary constraints $\chi^{0\sigma} = [ H_C, \phi^{0\sigma} ]$, where $\sigma = 0, 1, \ldots, d - 1$. 
This means that we have to add these $d$ non-zero secondary constraints $\chi^{0\sigma}$ to this Hamilton formulation \cite{constr}. The explicit 
formulas for the secondary constraints $\chi^{0\sigma}$ are very cumbersome and can be found in \cite{K&K} (see also \cite{Fro1}). Here we do not 
describe derivation of these and other similar formulas, since they were derived earlier in \cite{K&K}, and they are not original for this study. Our 
formulas for Poisson brackets substantially simplify the whole process of derivation of these formulas for the primary and secondary constraints and
for their commutators. In particular, by uising our Poisson brackets one can show that all Poisson brackets between primary constraints equal zero 
identically, i.e., $[ \phi^{0\lambda}, \phi^{0\sigma} ] = 0$, while $[ \phi^{0\lambda}, \chi^{0\sigma} ] = \frac12 g^{\lambda\sigma}$. The Poisson 
brackets between canonical Hamiltonian $H_C$ and secondary constraints $\chi^{0\sigma}$ are expressed as `quasi-linear' \cite{QL} combinations of the 
same secondary constrains $\chi^{0\sigma}$, i.e., we obtain
\begin{eqnarray}
 [ \chi^{0\sigma}, H_{c} ] &=& -\frac{2}{\sqrt{-g}} I_{mnpq} \pi^{mn} \Bigl(\frac{g^{\sigma q}}{g^{00}}\Bigr) \chi^{0p} + \frac12 g^{\sigma k} 
 g_{00,k} \chi^{00} + \delta_{0}^{\sigma} \chi_{,k}^{0k} \label{close} \\
 &+& \Bigl( -2 \frac{1}{\sqrt{-g}} I_{mnpk} \pi^{mn} \frac{g^{\sigma p}}{g^{00}} + I_{mkpq} g_{\mu\nu,l} \frac{g^{\sigma m}}{g^{00}} 
 A^{(pq) 0 \mu\nu l} \Bigr)\chi^{0k} \nonumber \\
 &-& \Bigl( g^{0\sigma} g_{00,k} + 2 g^{n\sigma} g_{0n,k} + \frac{g^{n\sigma} g^{0m}}{g^{00}} (g_{mn,k} + g_{km,n} - g_{kn,m}) \Bigr) \chi^{0k} 
 \nonumber
\end{eqnarray}
where $A^{(pq) 0 \mu\nu k}$ is the symmetrized form (upon all $p \leftrightarrow q$ permutations) of the following expression 
\begin{eqnarray}
 A^{pq 0 \mu\nu k}= B^{(pq 0 \mid \mu \nu k)} - g^{0k} E^{pq \mu \nu} + 2 g^{0\mu} E^{pq k \nu}.
\end{eqnarray}
The Poisson bracket, Eq.(\ref{close}), indicates that the Hamilton procedure developed for the metric GR in \cite{K&K} and \cite{FK&K} is closed 
(Dirac closure), i.e., the Poisson bracket $[ \chi^{0\sigma}, H_{c} ]$ does not lead to any tertiary, or other constraints of higher order(s). 
Analogously, the Poissonbrackets between secondary constraints $[ \chi^{0\sigma}, \chi^{0\gamma}]$, where $\sigma \ne \gamma$ (when $\sigma = 
\gamma$ this PB equals zero identically), is 
\begin{eqnarray}
 [ \chi^{0\sigma}, \chi^{0\gamma} ] &=& [ \chi^{0\sigma}, [ \phi^{0\gamma}, H_{c} ]] = - [ \phi^{0\gamma}, [ H_C, \chi^{0\sigma} ]] - 
 [ H_C, [ \chi^{0\sigma}, \phi^{0\gamma} ]] \nonumber \\ 
 &=& [ \phi^{0\gamma}, [ \chi^{0\sigma}, H_C ]] - \frac12 [ g^{\sigma\gamma}, H_C ] \; \; , \; \label{chichi} 
\end{eqnarray} 
where the Poisson bracket $[ \chi^{0\sigma}, H_C ]$ is given by the formula, Eq.(\ref{close}). This formula also does not lead to any constraint
of higher order and/or to any other expression which is not a function of the dynamical variables only (see dscussion in \cite{Dir64}). This proves 
that the Hamiltonian system which includes the canonical Hamiltonian $H_C$ and all primary $\phi^{0\lambda}$ and secondary $\chi^{0\sigma}$ 
constraints  \cite{constr} is closed (here $\lambda = 0, 1, \ldots, d - 1$ and $\sigma = 0, 1, \ldots, d - 1$). The actual closure of the Dirac 
procedure \cite{Dir50} for the Hamiltonian formulation of the metric GR considered above was shown for the first time in \cite{K&K}. Formally, the 
explicit demonstration of closure of the whole Dirac procedure \cite{Dir50} is the last and most important step for any Hamiltonian formulation of 
the metric GR \cite{Dir64}. However, in reality one needs to check one more condition which appears to be crucial for separation of the actual 
Hamiltonian formulations of the metric GR from numerous quasi-Hamiltonian constructions developed in this area, since the middle of 1950's. 

This additional condition is the rigorous conservation of the bothe true (or algebraic) and gauge symmetries of the metric GR which coincides with 
the symmetry of original Einstein's equation(s) for the free gravitational field. In general, by performing a chain of transformations from the 
original $\Gamma - \Gamma$ Lagrangian to the Hamiltonian formulation of the metric GR we have to be sure that all regular and gauge symmetries (or 
invariances) are conserved. Disappearance (or reduction) of the original gauge symmetry of the problem simply means that our transformations to the 
Hamiltonian formulation are fundamentally wrong, or simply that `they are not canonical'. Our formulas for the Hamiltonians $H_t, H_C$ presented 
above and explicit expressions for all primary and secondary constraints \cite{K&K}, \cite{Fro1} allow one to derive (with the use of Castellani 
procedure \cite{Cast}) the correct generators of gauge transformations, which directly and unambogously lead to the diffeomorphism invariance 
\cite{K&K}. This diffeomorphism invariance is well known gauge symmetry (or gauge, for short) for the free gravitational field(s) since early years 
of the metric GR (see, e.g., \cite{Carm}). Currently, there are only two known Hamiltonian formulations developed for the metric GR (\cite{Dir58} 
and \cite{K&K}) which are able to reproduce the actual diffeomorphism invariance directly and transparently. Note that for all approaches, which 
are based on the $\Gamma - \Gamma$ Lagrangian of the metric GR, such a reconstruction of the diffeomorphism invariance (or gauge) is a relatively 
simple problem (see, e.g., \cite{Saman}). In contrast with this, for any Hamiltonian-based formulation the complete solution of similar problem 
requires a substantial work. However, it is clear that analytical derivation of the diffeomorphism invariance is a very good test for the total 
$H_t$ and canonical $H_C$ Hamiltonians as well as for all primary $\phi^{0\sigma}$ and secondary $\chi^{0\sigma}$ constraints derived in any new 
Hamiltonian formulation of the metric GR. Any mistake either in the $H_t$and $H_C$ Hamiltonians, or in the $\phi^{0\lambda}$ and $\chi^{0\sigma}$ 
constraints leads to the loss of true diffeomorphism invariance.   
   
\subsection{Hamilton equations of motion for the free gravitational field}

In general, if we know the total $H_t$ and canonical $H_C$ Hamiltonians, Eqs.(\ref{eq1}) and (\ref{eq5}), respectively, then we can derive the 
Hamilton equations of motion (or system of Hamilton equations) which describe the time-evolution of all dynamical variables in the metric GR, 
i.e., time-evolution of each component of the metric tensor $g_{\alpha\beta}$ and momentum tensor $\pi^{\gamma\rho}$. These equations are 
\cite{Fro1}
\begin{eqnarray}
 \frac{d g_{\alpha\beta}}{d x_0} = [ g_{\alpha\beta}, H_{t} ] \; \; \; {\rm and} \; \; \;  \frac{d \pi^{\gamma\rho}}{d x_0} = [ \pi^{\gamma\rho}, 
 H_{t} ] \label{eq20}
\end{eqnarray}
where the notation $x_0$ denotes the temporal variable. In particular, for the spatial components $g_{ij}$ of the metric tensor $g_{\alpha\beta}$ one 
finds the following equations
\begin{eqnarray}
 \frac{d g_{ij}}{d x_0} &=& [ g_{ij}, H_{t} ] = [ g_{ij}, H_{c} ] = \frac{2}{\sqrt{-g} g^{00}} I_{(ij)pq} \pi^{pq} - \frac{1}{g^{00}} I_{(ij)pq} 
 B^{(p q 0|\mu \nu k)} g_{\mu\nu,k} \; \label{eq25} \\
 &=& \frac{2}{\sqrt{-g} g^{00}} I_{(ij)pq} \Bigl[ \pi^{pq} - \frac12 \sqrt{-g} B^{(p q 0|\mu \nu k)} g_{\mu\nu,k} \Bigr] \nonumber 
\end{eqnarray}
where the notation $I_{(ij)pq}$ stands for the $(ij)-$symmetrized values of the $I_{ijpq}$ tensor defined in Eq.(\ref{I}), i.e., 
\begin{equation}
 I_{(ij)pq} = \frac12 \Bigl( I_{ijpq} + I_{jipq} \Bigr) = \frac{1}{d - 2} g_{ij} g_{pq} - \frac12 ( g_{ip} g_{jq} + g_{iq} g_{jp} ) \; \; \; .
\end{equation}
Analogously, for the $g_{0\sigma}$ components of the metric tensor one finds the following equations of time-evolution
\begin{eqnarray}
 \frac{d g_{0\sigma}}{d x_0} = [ g_{0\sigma}, H_{t} ] = g_{0\sigma,0} \; , \; \; \label{eq253}
\end{eqnarray}
since all $g_{0\sigma}$ components commute with the canonical Hamiltonian $H_C$, Eq.(\ref{eq5}), while all $g_{ij}$ commute with the primary 
constraints $\phi^{0\sigma}$. This result could be expected, since the equation, Eq.(\ref{eq253}), is, in fact, a definition of the 
$\sigma-$velocities (or $g_{0\sigma,0}$-velocities), where $\sigma = 0, 1, \ldots, d - 1$.   

The Hamilton equations for tensor components of the momentum $\pi^{\alpha\beta}$, Eq.(\ref{eq20}), are substantially more complicated. They 
are derived by calculating the Poisson brackets between each term in $H_{t}$ and $\pi^{\gamma\rho}$. This general formula takes the form
\begin{eqnarray}
 \frac{d \pi^{\alpha\beta}}{d x_0} &=& - [ H_{t}, \pi^{\alpha\beta} ] = - \Bigl[ \frac{I_{mnpq}}{\sqrt{-g} g^{00}}, \pi^{\alpha\beta} \Bigr] 
 \pi^{mn} \pi^{pq} \nonumber \\
 &+& \Bigl[ \frac{I_{mnpq}}{g^{00}}, \pi^{\alpha\beta} \Bigr] \pi^{mn} B^{(p q 0|\mu \nu k)} g_{\mu\nu,k} 
  + \frac{1}{g^{00}} I_{mnpq} \pi^{mn}\Bigl[ B^{(p q 0|\mu \nu k)}, \pi^{\alpha\beta} \Bigr] g_{\mu\nu,k} + \ldots \; \label{eq255}
\end{eqnarray}
Let us determine the first Poisson bracket in this formula (other terms are considered analogously, i.e., term-by-term). The explicit expression 
for this term is 
\begin{eqnarray}
 &-& \Bigl[ \frac{I_{mnpq}}{\sqrt{-g} g^{00}}, \pi^{\alpha\beta} \Bigr] \pi^{mn} \pi^{pq} = - \frac{[ I_{mnpq}, \pi^{\alpha\beta}]}{\sqrt{-g} 
 g^{00}} \pi^{mn} \pi^{pq} - [ \frac{1}{\sqrt{-g} g^{00}}, \pi^{\alpha\beta} \Bigr] I_{mnpq} \pi^{mn} \pi^{pq} \; \; \label{eq256}
\end{eqnarray}
There are three following cases: (1) for a pair of space-like indexes, i.e., for $(\alpha\beta) = (a b)$, we have  
\begin{eqnarray}
 \Bigl( \frac{d \pi^{a b}}{d x_0}\Bigr)_1 = -\frac{2}{d - 2} g_{m n} \pi^{m n} \pi^{a b} + 2 g_{m p} \pi^{m a} \pi^{p b} + 
 \frac{I_{mnpq}}{2 \sqrt{-g} g^{00}} g^{a b} \pi^{m n} \pi^{p q} \; \; \; \label{eq257}
\end{eqnarray}
while for the $(\alpha\beta) = (0 a)$ indexes the expression is 
\begin{eqnarray}
 \Bigl( \frac{d \pi^{0 a}}{d x_0}\Bigr)_1 = \frac{I_{mnpq}}{2 \sqrt{-g} g^{00}} g^{0 a} \pi^{m n} \pi^{p q} \; \; \label{eq2561}
\end{eqnarray}
Finally, for the $(\alpha\beta) = (0 0)$ pair of indexes one finds
\begin{eqnarray}
 \Bigl( \frac{d \pi^{0 0}}{d x_0}\Bigr)_1 = \frac{I_{mnpq}}{2 \sqrt{-g}} \Bigl( 1 + \frac{2}{(g^{00})^{2}} \Bigr) \pi^{mn} \pi^{pq} \; \; 
 \label{eq2562}
\end{eqnarray}
In general, analytical calculations of other Poisson brackets in the formula, Eq.(\ref{eq255}), is a straightforward task, but the final formula 
contains more than 150 terms. This drastically complicates all operations with the formula, Eq.(\ref{eq255}), for the $\frac{d \pi^{\gamma\rho}}{d 
x_0}$ (temporal) derivative. Nevertheless, the complete set of Hamilton equations for the free gravitational field in metric GR has been produced 
in closed and explicit form \cite{FroUnp}. 

\subsection{Truly canonical transformations in the metric GR}

As is well known all canonical transformations for an arbitrary Hamilton system form a closed algebraic group. This means that in any Hamilton 
system: (1) consequence of the two canonical transformations is the new canonical transformation, (2) identical transformation of dynamical 
variables is the canonical transformation, (3) any canonical transformation has its inverse transformation which is also canonical and unique. 
In general, there are quite a few canonical transformations in the metric General Relativity, and some of them can be used to simplify either 
Hamiltonian(s), or secondary constraints, or some other crucial quantities, including a few important Poisson brackets. As is well known (see, 
e.e., \cite{LLTF}, \cite{Carm}) the metric General Relativity is a non-linear theory which cannot rigorously be linearized even in lower-order 
approximations. Therefore, the linear canonical transformations of dynamical variables have no interest for the Hamiltonian formulations which 
have been developed for the metric GR. Furthermore, it can be shown that among all possible non-linear canonical transformations the following 
`special' transformations play a great role in derivation of the new Hamiltonian formulations of the metric GR. These special canonical 
transformations can be written in the form: $\{ g_{\alpha\beta}, \pi^{\mu\nu}\} \rightarrow \{ g_{\alpha\beta}, \Pi^{\rho\sigma}\}$, where the 
new momenta $\Pi^{\rho\sigma}$ are the linear functions (or linear combinations) of old momenta $\pi^{\mu\nu}$ to which a cubic functions (or
cubic polynomials) of the contravariant components of metric tensor $g^{\alpha\beta}$. The coefficient(s) in front of this cubic function can 
also contain factors such as $\sqrt{- g}$ and/or $g^{00}$, or their product. As follows from our experience only such canonical transformations 
can be used for equivalent transformation of the two different sets of dynamical variables in the metric GR. In particular, this form can be 
found for the canonical transformation of dynamical variables constructed in \cite{FK&K} has such a form. This canonical transformation relates 
the two correct Hamiltonian formulations known to this moment in metric GR, i.e., formulation by Dirac \cite{Dir58} and K$\&$K \cite{K&K} 
formulation. Our new canonical transformation of dynamical variables described below also has this form. Furthermore, if some `new' set of 
dynamical variables (in metric GR) is related to another `old' set of dynamical variables by a canonical transformation which has the mentioned 
form, then it can be shown that this transformation of variables will preserve the complete diffeomorphism as a gauge symmetry of the free 
gravitational field. Very likely, the explicit form ofsuch `special' canonical transformations and all possible consequencies of this fact are 
substantially determined by the $\Gamma - \Gamma$ Lagrangian presented in Section II. Indeed, the $\Gamma - \Gamma$ Lagrangian, Eq.(\ref{eq05}), 
is a polinomial of power six upon the $g^{\alpha\beta}$ components and a quadratic function of the space-like velocities $g_{mn,0}$.

\section{Canonical Hamiltonian reduced to its natural form}

In this Section we reduce the canonical Hamiltonian $H_C$ to its natural form, which will play a significant role in numerous applications to 
the metric gravity. We perform such a reduction of $H_C$ by using some canonical transformation of the dynamical variables $g_{\alpha\beta}$ 
and $\pi^{\rho\sigma}$ defined above. First, let us write the canonical Hamiltonian, Eq.(\ref{eq5}), in the form
\begin{eqnarray}
 H_C &=& \frac{I_{mnpq}}{\sqrt{-g} g^{00}} \Bigl[ \pi^{mn} \pi^{pq} - \sqrt{-g} \pi^{mn} B^{(p q 0|\mu \nu k)} g_{\mu\nu,k} +
 \frac14 (- g) B^{(m n 0|\mu \nu k)} B^{(p q 0|\alpha \beta l)} g_{\mu\nu,k} g_{\alpha\beta,l} \Bigr] \nonumber \\
 &+& \frac14 \sqrt{-g} \Bigl \{ \frac{1}{g^{00}} I_{mnpq} B^{([mn] 0|\mu\nu k)} B^{(p q 0|\alpha \beta l)} - 
 B^{\mu\nu k \alpha\beta l}\Bigr \} g_{\mu\nu,k} g_{\alpha\beta,l} \; \; \label{eq5a} 
\end{eqnarray}
which is more appropriate for our purposes in this study. In Eq.(\ref{eq5a}) the notation $B^{([mn] 0|\mu\nu k)}$ stands for the $B^{(m n 0 
\mid \mu\nu k)}$ cubic function of the contravariant components of the metric tensor which is completely anti-symmetric in respect to the 
$m$ and $n$ indexes. The explicit formula for the $B^{([mn] 0|\mu\nu k)}$ function is 
\begin{eqnarray}
 B^{([mn] 0|\mu\nu k)} &=& g^{m k} g^{n \nu} g^{\nu 0} - g^{n k} g^{m \nu} g^{\nu 0} + \frac12 \Bigl( g^{n \mu} g^{m \nu} g^{k 0} + 
 g^{n k} g^{\mu \nu} g^{m 0} - g^{m \mu} g^{n \nu} g^{k 0} \nonumber \\ 
 &-& g^{m k} g^{\mu \nu} g^{m 0} \Bigr) \; \; \label{AsBcoef}
\end{eqnarray}
Now, we can see that the first term in $\Bigl[ \ldots \Bigr]$ brackets in Eq.(\ref{eq5a}) can be written as a pure quadratic function of the new 
$P^{mn} = \pi^{mn} - \frac12 \sqrt{-g} B^{(m n 0|\mu\nu k)} g_{\mu\nu, k}$ variables, i.e., 
\begin{eqnarray}
 H_C &=& \frac{I_{mnpq}}{\sqrt{-g} g^{00}} \Bigl( \pi^{mn} - \frac12 \sqrt{-g} B^{(m n 0|\mu\nu k)} g_{\mu\nu, k} \Bigr) 
 \Bigl( \pi^{pq} - \frac12 \sqrt{-g} B^{(p q 0|\alpha\beta l)} g_{\alpha\beta, l} \Bigr) \nonumber \\
 &+& \frac14 \sqrt{-g} \Bigl\{ \frac{1}{g^{00}} I_{mnpq} B^{([mn] 0|\mu\nu k)} B^{(p q 0|\alpha \beta l)} - B^{\mu\nu k \alpha\beta l}\Bigr\} 
 g_{\mu\nu,k} g_{\alpha\beta,l} + T_1 + T_2 \; \; , \; \label{H_Cnew}
\end{eqnarray}
where the two additional terms $T_1$ and $T_2$ take the following form 
\begin{eqnarray}
 T_1 = \frac{I_{mnpq}}{2 \sqrt{-g} g^{00}} [ \pi^{mn}, \sqrt{- g}] B^{(p q 0|\alpha \beta l)} g_{\alpha\beta,l} = - \frac{I_{mnpq} g^{mn}}{2 
 g^{00}} B^{(p q 0|\alpha \beta l)} g_{\alpha\beta,l} \; \; \; \label{eq5b} 
\end{eqnarray}
and 
\begin{eqnarray}
 T_2 &=& - \frac{I_{mnpq}}{2 g^{00}} [ B^{(m n 0|\mu \nu k)}, \pi^{pq} ] g_{\mu\nu,k} = - \frac{I_{mnpq}}{2 g^{00}} \Bigl[ \frac12 \Bigl( 
 g^{\mu p} g^{m q} + g^{\mu q} g^{m p} \Bigr) g^{n \nu} g^{k 0} \nonumber \\
 &+& \frac12 g^{\mu m} \Bigl( g^{n p} g^{\nu q} + g^{n q} g^{\nu p} \Bigr) g^{k 0} + \frac12 g^{\mu m} g^{n \nu} \Bigl( g^{k p} g^{0 q} 
 + g^{k q} g^{0 p} \Bigr) \nonumber \\
 &-& \frac12 \Bigl( g^{m p} g^{n q} + g^{m p} g^{n q} \Bigr) g^{k 0} g^{\mu\nu} - \frac12 g^{m n} \Bigl( g^{p k} g^{q 0} + g^{p 0} 
   g^{q k} \Bigr) - \frac12 g^{m n}  g^{k 0} \Bigl( g^{\mu p} g^{\nu q} + g^{\mu q} g^{\nu p} \Bigr) \nonumber \\ 
 &-& \Bigl( g^{m p} g^{k q} + g^{m q} g^{k p} \Bigr) g^{n \nu} g^{\mu 0} - g^{m k} \Bigl( g^{n p} g^{\nu q} + g^{n q} g^{\nu p} \Bigr) 
    g^{\mu 0} - \frac12 g^{m k} g^{n \nu} \Bigl( g^{\mu p} g^{0 q} + g^{\mu q} g^{0 p} \Bigr)  \nonumber \\
 &+& \frac12 \Bigl( g^{m p} g^{n q} + g^{m q} g^{n p} \Bigr) g^{\nu k} g^{0 \mu} + \frac12 g^{m n} \Bigl( g^{\nu p} g^{k q} + g^{\nu q} 
 g^{k p} \Bigr) g^{\mu 0} + \frac12 g^{m n} g^{\nu k} \Bigl( g^{p 0} g^{\mu q} + g^{0 q} g^{\mu p} \Bigr) \nonumber \\
 &+& \frac12 \Bigl( g^{k p} g^{m q} + g^{k q} g^{m p} \Bigr) g^{\nu k} g^{0 \mu} + \frac12 g^{k m} \Bigl( g^{\mu p} g^{\nu q} + g^{p \nu} 
 g^{\mu q} \Bigr) g^{n 0} + \frac12 g^{k m} g^{\mu \nu} \Bigl( g^{n p} g^{0 q} \nonumber \\
 &+& g^{n q} g^{0 p} \Bigr) \Bigr] g_{\mu\nu,k} \; \; . \; \label{eq5c} 
\end{eqnarray}

Now, we can introduce the new momenta $P^{\gamma\rho}$ which is written in the form 
\begin{eqnarray}
 P^{\gamma\rho} = \pi^{\gamma\rho} - \frac12 \sqrt{-g} B^{(\gamma\rho 0|\mu\nu k)} g_{\mu\nu, k} \; \; \; \label{canvar}
\end{eqnarray}
where $\pi^{\gamma\rho}$ are the `old' momenta used in \cite{K&K}. These new momenta can be considered as the contravariant components of the 
tensor of `united' momentum $P = g_{\alpha\beta} P^{\alpha\beta}$. Note that the explicit expressions for the old velocities written in terms 
of new momenta $P^{ab}$ are even simpler $g_{mn, 0} = \frac{1}{\sqrt{-g} g^{00}} I_{m n q p} P^{pq}$ (compare with Eq.(\ref{veloc}) from above). 
The explicit formulas for the primary constraints are also simpler: $P^{0\gamma} \approx 0$ for $\gamma = 0, 1, \ldots, d - 1$. The generalized 
coordinates are chosen in the old (or traditional) form, i.e., they coincide with the covariant components of the metric tensor $g_{\alpha\beta}$. 
It is clear that similar choice of the generalized coordinates provides a number of additional advantages in applications to the metric GR. For 
instance, by using the metric tensor one can rise and lower indexes in arbitrary vectors and tensors. Also, all covariant and contravariant 
derivatives of the metric tensor always equal zero, i.e., this tensor behaves as a constant during these operations. More unique and remarkable 
properties of the metric tensor are discussed, e.g., in \cite{Kochin}. For the purposes of this study it is important to note only that our new 
system of dynamical variables contains the same `coordinates' $g_{\alpha\beta}$ and new momenta $P^{\gamma\rho}$. The Poisson brackets between 
our new dynamical variables can easily be determined by using the known values of Poisson brackets written in the old dynamical variables 
$\Bigl\{ g_{\alpha\beta}, \pi^{\gamma\rho} \Bigr\}$ defined above. We have $[ g_{\alpha\beta}, P^{\gamma\rho} ] = [ g_{\alpha\beta}, 
\pi^{\gamma\rho} ] = \Delta^{\gamma\rho}_{\alpha\beta} = \frac12 \Bigl( \delta^{\gamma}_{\alpha} \delta^{\sigma}_{\beta} + 
\delta^{\sigma}_{\alpha} \delta^{\gamma}_{\beta} \Bigr), [ g_{\alpha\beta}, g_{\gamma\rho} ] = 0$ (these two basic variables coincide with 
the original (or traditional) `coordinates' used in \cite{Dir58}, \cite{K&K}, \cite{FK&K}) and $[ P^{\alpha\beta}, P^{\gamma\rho} ] = 0$. The 
last equality we consider in detail
\begin{eqnarray}
 & &[ P^{\alpha\beta}, P^{\gamma\rho} ] = [ \pi^{\alpha\beta}, \pi^{\gamma\rho} ] - \frac12 [ \sqrt{-g} B^{(\alpha\beta 0|\mu\nu k)}, 
 \pi^{\gamma\rho} ] g_{\mu\nu, k} + \frac12 [ \sqrt{-g} B^{(\alpha\beta 0|\lambda\sigma, l)}, \pi^{\gamma\rho} ] g_{\lambda\sigma, l} 
 \nonumber \\
 &+& [ \sqrt{-g} B^{(\alpha\beta 0|\mu\nu k)} g_{\mu\nu, k}, \sqrt{-g} B^{(\gamma\rho 0|\lambda\sigma l)} g_{\lambda\sigma, l} ] \label{PBV}
\end{eqnarray}
where the first and last terms equal zero identically, since the variables $g_{\alpha\beta}$ and $\pi^{\mu\nu}$ are canonical. This directly 
leads to the formula 
\begin{eqnarray}
 [ P^{\alpha\beta}, P^{\gamma\rho} ] = - \frac12 [ \sqrt{-g} B^{(\alpha\beta 0|\mu\nu k)}, \pi^{\gamma\rho} ] g_{\mu\nu, k} + 
 \frac12 [ \sqrt{-g} B^{(\alpha\beta 0|\lambda\sigma, l)}, \pi^{\gamma\rho} ] g_{\lambda\sigma, l}  \label{PBV1}
\end{eqnarray}
Now, we can replace the dummy indexes in the second term of this equation by the values which coincide with the corresponding dummy indexes 
in the first term,i.e., $\lambda \rightarrow \mu, \sigma \rightarrow \nu$ and $l \rightarrow k$. This substitution reduces Eq.(\ref{PBV1}) 
to the form 
\begin{eqnarray}
 [ P^{\alpha\beta}, P^{\gamma\rho} ] = - \frac12 [ \sqrt{-g} B^{(\alpha\beta 0|\mu\nu k)}, \pi^{\gamma\rho} ] g_{\mu\nu, k} + 
 \frac12 [ \sqrt{-g} B^{(\alpha\beta 0|\mu\nu, k)}, \pi^{\gamma\rho} ] g_{\mu\nu, k} = 0 \label{PBV2}
\end{eqnarray}
which is the difference of the two identical expressions. This shows that the new dynamical variables $\{ g_{\alpha\beta}, P^{\mu\nu}\}$ are 
also canonical, and they can be used in the metric gravity, since they are canonically related to the old set of such variables 
$\{ g_{\alpha\beta}, \pi^{\mu\nu}\}$ \cite{K&K}. 

As follows from the formulas derived above the canonical Hamiltonian $H_C$ is reduced to the following form 
\begin{eqnarray}
 H_C &=& \frac{I_{mnpq}}{\sqrt{-g} g^{00}} P^{mn} P^{pq} + \frac14 \sqrt{-g} \Bigl[ \frac{I_{mnpq}}{g^{00}} B^{([mn] 0|\mu\nu k)} 
 B^{(p q 0|\alpha \beta l)} - B^{\mu\nu k \alpha\beta l}\Bigr] g_{\mu\nu,k} g_{\alpha\beta,l} \nonumber \\ 
 &-& \frac{I_{mnpq}}{2 g^{00}} g^{mn} B^{(pq 0| \alpha\beta l)} g_{\alpha\beta,l} + T_2 \; \; \label{eq5d} 
\end{eqnarray}
which can be re-written in the following symbolic form 
\begin{eqnarray}
  H_C = \frac12 \sum^{n}_{i,j=1} \hat{M}_{ij}(q_1, q_2, \ldots, q_n) p_i p_j + \sum^{n}_{i,j=1} \hat{V}_{mn}(q_1, q_2, \ldots, q_n) 
  \; \; \label{ClassH} 
\end{eqnarray}
where $\hat{M}$ is a positively defined $n \times n$ matrix which is often called the inverse mass matrix (or matrix of inverse masses), 
while the $\hat{V}$ matrix is an arbitrary, in principle, symmetric $n \times n$ matrix which is called the potential matrix (or matrix 
of the potential energy). Here $n$ is the total number of generalized coordinates $q_1, q_2, \ldots, q_n$. Each matrix element of the 
potential matrix $\hat{V}$ in Eq.(\ref{ClassH}) is a polynomial of these generalized coordinates. Also, in Eq.(\ref{ClassH}) the notations 
$p_i$ and $p_j$ designate the momenta conjugate to the corresponding generalized coordinates $q_i$ and $q_j$, respectively, i.e., $[ q_k, 
p_l] = \delta_{kl}$. In classical mechanics the phase space is flat, and, therefore, the both covariant and contravariant components of 
any vector coincide with each other. The form of the Hamiltonian $H_C$, Eq.(\ref{ClassH}), is called normal, and it is well known in 
classical mechanics of Hamiltonian systems. Furthermore, more than 90 \% of all problems ever solved in classical Hamiltonian mechanics 
with the use of Hamilton methods either have Hamiltonians which are already written in the normal form, or their Hamiltonians can be 
reduced to such a form by some canonical transformation(s) of variables. 
 
To improve the overall quality of our analogy between metric GR and classical Hamiltonian mechanics one can introduce the new set of 
dynamical variables which include the total momentum of the free gravitational field $P = g_{\alpha\beta} P^{\alpha\beta}$ (tensor 
invariant) and its tensor `projections' $P_{\alpha}^{\beta} = g_{\alpha\gamma} P^{\gamma\beta}$. The corresponding space-like quantities 
$P = g_{mn} P^{mn}$ and $P_{m}^{n} = g_{m p} P^{p n}$ are already included in our canonical Hamiltonian $H_C$. By using our formulas 
presented above one easily finds a few following Poisson brackets:
\begin{eqnarray}  
 &[& P, P^{ab} ] = [ g_{mn}, P^{ab} ] P^{mn} = \Delta^{ab}_{mn} P^{mn} = P^{ab} \; , \; [ g_{cd}, P ] = g_{mn} [ g_{cd}, P^{mn} ] 
 = g_{cd} \nonumber \\
 &[& g_{\alpha\beta}, P^{\gamma}_{\sigma} ] = \frac12 ( g_{\beta\sigma} \delta^{\gamma}_{\alpha} + g_{\alpha\sigma} 
 \delta^{\gamma}_{\beta} ) \; \; , \;  [ g^{\alpha\beta}, P ] = g^{\alpha\beta} \; \; \; 
 \nonumber 
\end{eqnarray}
and many others. Here we cannot present all of them explicitly. Note only that with the total momentum $P$ and its tensor projections
(i.e., $P^{\alpha\beta}, P^{\gamma}_{\sigma}$, etc) one can write the Hamilton equations in the form which is almost coincides with 
analogous equations known for Hamiltonian systems in classical mechanics. This is another interesting direction for future development 
of the Hamiltonian formulation(s) of metric GR. Applications of our new canonical variables $\{ g_{\lambda\kappa}, P^{\alpha\beta} \}$
to some interesting problems in metric GR  will be considered elsewhere. Relations between our dynamical variables $\{ g_{\lambda\kappa}, 
P^{\alpha\beta} \}$ and analogous variables used in Dirac formulation of the metric General Relativity $\{ g_{\lambda\kappa}, 
\pi^{\alpha\beta} \}$ are discussed in the Appendix A. 

\section{Discussions and Conclusion}

Thus, we have shown that the canonical Hamiltonian $H_C$ of the free gravitational field(s), Eq.(\ref{eq5a}), can be reduced to the natural 
form which includes a pure quadratic function of the space-like momenta $P^{mn}$ with a positive coefficient in front of it. Indeed, the 
factor, which is located in front of the $P^{mn} P^{pq}$ product in the $H_C$ Hamiltonian, is the positively defined space-like tensor of 
the fourth rank $I_{mn pq}$ (or $\frac{1}{\sqrt{-g}} I_{mn pq}$). This factor can be considered as an effective inverse `quasi-mass' of the 
free gravitational field in metric GR. Also, as directly follows from the explicit form of the canonical Hamiltonian $H_C$, Eq.(\ref{eq5a}), 
each of the remaining terms in canonical Hamiltonian $H_C$, Eq.(\ref{eq5a}), is a polynomial function of contravariant components 
$g^{\alpha\beta}$ of the metric tensor. The maximal power of such polynomials upon $g^{\alpha\beta}$ does not exceed eight. Some terms in 
the $H_C$ also include the factors $\sqrt{-g}$ (or $\frac{1}{\sqrt{-g}})$ and/or $g^{00}$.   

The new canonical $\{ g_{\alpha\beta}, P^{\gamma\rho} \}$ variables have been constructed for the metric GR. The total number of canonical 
variables does not changed and it always equals $2 d$. The Poisson brackets between these variables are: $[ g_{\alpha\beta}, 
P^{\gamma\rho} ] = \Delta^{\gamma\rho}_{\alpha\beta} = \frac12 \Bigl( \delta^{\gamma}_{\alpha} \delta^{\rho}_{\beta} + \delta^{\rho}_{\alpha} 
\delta^{\gamma}_{\beta} \Bigr) = [ P_{\gamma\rho}, g^{\alpha\beta} ], [ g_{\alpha\beta}, g_{\gamma\sigma} ] = 0$and $[ P^{\alpha\beta}, 
P_{\gamma\rho} ] = 0$. This indicates clearly that these new dynamical variables are truly canonical and can be used in the new Hamiltonian 
formulation of the General Relativity. Analogous set of dynamical variables $\{ g^{\alpha\beta}, P_{\gamma\rho} \}$ is the dual set of 
canonical variables which can also be used to develop a different (but equivalent!) Hamiltonian formulation of the metric GR. 

Thus, in this study we have finished development of the complete and correct Hamiltonian formulation of the metric General Relativity. Also, 
we have determined all essential (fundamental and secondary) Poisson brackets which can now be used to perform a large amount of analytical 
and numerical calculations. The fundamental Poisson brackets are defined between all components of the gravitational fieldand corresponding 
momenta (or components of the momentum tensor). The secondary Poisson brackets define commutation relations between arbitrary, in principle, 
analytical functions of coordinates (components of the gravitational field) and momenta. These Poisson brackets become the main working tools 
of the metric General Relativity, which can now be considered as a Hamiltonian system. In addition to this, our Poisson brackets can be used 
to solve various problems in metric GR, e.g., obtain trajectories, derive conservation laws, find integrals of motion, derive and investigate 
the laws of time-evolution for different quantities, vectors and tensors. A remarkable result obtained in this study should be emphasized 
again: the canonical Hamiltonian $H_C$, which describes time-evolution of relativistic gravitational fields, can be reduced to its natural 
form, and this form essentially coincides with the Hamiltonian of the non-relativistic system of $N (= d)$ interacting particles. Physical 
sense of dynamical variables is obviously very different in both these cases, but almost identical coincidence of their Hamiltonians was 
absolutely unexpected and shocking. 

In conclusion, it should be emphasized again that the first non-cotradictory Hamiltonian formulation of metric GR was presented by P.A.M. 
Dirac in 1958 \cite{Dir58}. The second `alternative' formulation was developed in \cite{K&K}. The both these correct Hamiltonian 
formulations of metric GR preserve the complete diffeomorphism as the gauge symmetry of this theory. In our earlier paper \cite{FK&K} we 
have shown that these two Hamiltonian formulations are related by a true canonical transformation $\{ g^{\alpha\beta}, \pi^{\gamma\rho} \} 
\rightarrow \{ g^{\alpha\beta}, p^{\gamma\rho} \}$. In this study we have solved a number of remaining problems which were never discussed 
in earlier papers. In particular, we obtained formulas for various Poisson brackets which are need in different Hamiltonian formulation(s) 
of the metric GR. This also includes the Poisson brackets from the two sets of basic dynamical variables: (a) set of straight (or Dirac) 
dynamical variables, e.g., $\{ g_{\alpha\beta}, \pi^{\gamma\rho} \}$ (or $\{ g_{\alpha\beta}, P^{\gamma\rho} \}$), and (b) dual set of basic 
dynamical variables $\{ g^{\alpha\beta}, \pi_{\gamma\rho} \}$ (or $\{ g^{\alpha\beta}, P_{\gamma\rho} \}$). The fundamental relation between 
these two sets of dynamical variables is given by the Poisson bracket, Eq.(\ref{eq1551}). In our new dynamical variables the same relation
takes the form $[ g_{\alpha\beta}, P^{\mu\nu}] = \Delta^{\mu\nu}_{\alpha\beta} = [ P_{\alpha\beta}, g^{\mu\nu}]$. Applications of our 
Hamiltonian formulation of the metric GR to some interesting problems will be considered in the next studies.  

Finally, as we all know many physists called and considered the General Relativity (or metric GR in our words) as "the most beautiful 
of all existing physical theories" (see, e.g., \cite{LLTF}, page 228). Here I wish to note that the correct Hamiltonian formulation 
of the metric General Relativity (or, Gravity, for short) is also very beautiful physical theory. Furthermore, the truly covariant, very 
powerful and explicitly beautiful apparatus of this theory corrects everybody (even authors), if they steps away from the unique, truly 
covariant and correct road of actual theory. No comparison can be made with an ugly form of the original geometro-dynamics (see, Appendix B) 
and similar Hamiltonian-like creations, which were decleared to be `canonicaly related' with the geometro-dynamics. 
    
I am grateful to my friends N. Kiriushcheva, S.V. Kuzmin and D.G.C. (Gerry) McKeon (all from the University of Western Ontario, London, 
Ontario, Canada) for helpful discussions and inspiration.

{\bf Appendix A}

In this Appendix we discuss relations between dynamical variables which are used in our and Dirac formulations of the metric General Relativity. 
In our earlier papers \cite{FK&K} we have shown that dynamical variables $\{ g_{\lambda\kappa}, \pi^{\alpha\beta} \}$, which are used in the $K\&K$
formulation of the metric GR, and analogous Dirac dynamical variables $\{ g_{\lambda\kappa}, p^{\alpha\beta} \}$ of the metric GR \cite{Dir58} are 
related to each other by some canonical transfromation. This canonical transfromation can be written in the form \cite{FK&K} (from Dirac to $K\&K$)
\begin{eqnarray}
  g_{\lambda\kappa} \rightarrow g_{\lambda\kappa} \; \; \; {\rm and} \; \; \; p^{\alpha\beta} \rightarrow \pi^{\alpha\beta} - \frac12 \sqrt{- g} 
 A^{(\alpha\beta) 0 \mu \nu k} g_{\mu\nu,k} \; \; \label{p_mom}
\end{eqnarray}
where the quantity $A^{(\alpha\beta) 0 \mu \nu k}$ is 
\begin{eqnarray}
  A^{(\alpha\beta) 0 \mu \nu k} = B^{((\alpha\beta) 0 \mid \mu \nu k)} - g^{0 k} E^{(\alpha\beta) \mu\nu} + 2 g^{0 \mu} E^{(\alpha\beta) k\nu}
\end{eqnarray}
where $B^{((\alpha\beta) 0 \mid \mu \nu k)}$ is the $B^{(\alpha\beta 0 \mid \mu \nu k)}$ quantity (see, Eq.(\ref{Bcoef})) symmetrized in terms of 
all $\alpha \leftrightarrow \beta$ permutations. Analogously, the $E^{(\alpha\beta) \mu\nu}$ and $E^{(\alpha\beta) k\nu}$ are the two symmetrized 
quantities (in respect to the $\alpha \leftrightarrow \beta$ permutations), i.e., 
\begin{eqnarray}
 E^{(\alpha\beta) \mu\nu} = e^{\alpha\beta} e^{\mu\nu} - \frac12 ( e^{\alpha\mu} e^{\beta\nu} + e^{\alpha\nu} e^{\beta\mu} ) \; \; {\rm and} \; 
 E^{(\alpha\beta) k\nu} = e^{\alpha\beta} e^{k\nu} - \frac12 ( e^{\alpha k} e^{\beta\nu} + e^{\alpha\nu} e^{\beta k} ) \nonumber
\end{eqnarray}
respectively. 

As is shown in the main text the relation between our dynamical variables and dynamical variables inroduced in \cite{K&K} is $g_{\lambda\kappa} 
\rightarrow g_{\lambda\kappa}$ and $P^{\alpha\beta} \rightarrow \pi^{\alpha\beta}$, where 
\begin{eqnarray}
 P^{\alpha\beta} \rightarrow \pi^{\alpha\beta} - \frac12 \sqrt{- g} B^{(\alpha\beta 0 \mid \mu \nu k)} g_{\mu\nu,k} \; \; \; 
\end{eqnarray}
From the last equation it is easy to obtain the following expression for our momenta $P^{\alpha\beta}$ written in terms of the Dirac momenta 
$p^{\alpha\beta}$  
\begin{eqnarray}
  P^{\alpha\beta} = p^{\alpha\beta} - \frac12 \sqrt{- g} \Bigl[ B^{([\alpha\beta] 0 \mid \mu \nu k)} - g^{0 k} E^{(\alpha\beta) 
 \mu\nu} + 2 g^{0 \mu} E^{(\alpha\beta) k\nu} \Bigr] \; \; \label{P_p}
\end{eqnarray} 
where the quantity $B^{([\alpha\beta] 0 \mid \mu \nu k)}$ is the $B^{(\alpha\beta 0 \mid \mu \nu k)}$ coefficient, Eq.(\ref{Bcoef}), 
anti-symmetrized in respect to all permutations of the $\alpha$ and $\beta$ indexes. The transformation of dynamical variales $g_{\lambda\kappa} 
\rightarrow g_{\lambda\kappa}$ and $P^{\alpha\beta} \rightarrow p^{\alpha\beta}$, Eq.(\ref{P_p}), is the canonical transformation (this can be 
shown in the same way as it is done in the main text (see also \cite{K&K}). Its inverse transformation is also canonical. This means that 
currently we have three different sets of dynamical variables which can be applied for the known and new Hamiltonian formulations of the metric 
GR: (a) Dirac variables, (b) $K\&K$ variables \cite{K&K}, and (c) our variables defined in this study. These three sets of dynamical variables 
are related to each other by simple canonical transformations. \\

{\bf Appendix B}

In this Appendix we want to show that dynamical variables which are used in geometro-dynamics \cite{ADM} are not canonical. Therefore, this theory 
has nothing to do with the regular Hamiltonian formulation(s) of the metric GR. Furthermore, this theory (geometro-dynamics) cannot canonicaly be 
related to any of the correct Hamiltonian formulations known for the metric GR. On the other hand, all similar `theories' which are canonicaly 
related to the geometro-dynamics are equaly wrong quasi-Hamiltonian constructions which cannot help anybody to solve problems currently known and 
constantly arising in the metric GR. 

The history of creation of geometro-dynamics, which is also often called the ADM gravity, is straightforward. After an obvious success of Dirac 
paper \cite{Dir58} a small group of young authors, which included Arnowitt, Deser and Misner \cite{ADM} (under general supervision of J.A. Wheeler), 
decided to create some alternative (but Dirac-like!) formulation of the metric GR. Dynamical variables in this ADM approach were chosen as follows. 
The generalized six coordinates coincide with the corresponding space-space components $g_{pq}$ of the metric tensor $g_{\alpha\beta}$ defined 
in the four-dimensional space-time (or (3+1)-dimensional space-time, if we want to be historically precise). Four remaining coordinates were chosen 
in the form: the "lapse" $N = \frac{1}{\sqrt{- g^{00}}}$ and three "shifts" $N^{k} = - \frac{g^{0k}}{g^{00}}$, where $k = 1, 2, 3$ (very likely, the 
idea to use these four coordinates was proposed by Wheeler). The corresponding momenta $\Pi^{mn}$ were simply taken from Dirac paper \cite{Dir58} 
(see also our Appendix A), i.e., they coincide with the $p^{mn}$ momenta introduced by Dirac (see Appendix A). The four remaining momenta were not 
defined in the original ADM papers. Probably, this was done, since these four momenta lead to the (primary) constraints anyway. In general, it is 
very hard to describe and discuss the internal logic of this quasi-theory, but we have to note that geometro-dynamics was carefully analyzed earlier 
in \cite{KK2011} with a large number of details and references. 

In fact, we do not need to bother ourselves with deep discussion of ADM formulation, since we already have their ten generalized coordinates (one 
laps $N$, three shifts $N^{k}$ and six components of the metric tensor $g_{pq}$) and six momenta $\Pi^{mn}$ which coincide with the momenta $p^{mn}$ 
defined in Dirac's paper. By using only these dynamic variables of ADM gravity we can prove that these variables are not canonical. To prove this 
statement we need to calculate the two following Poisson brackets: (1) between "laps" $N$ and $\Pi^{mn}$ (or $p^{mn}$) momenta, and (2) between 
"shifts" and the same $\Pi^{mn}$ (or $p^{mn}$) momenta. If this theory is a truly Hamiltonian, then all these Poisson brackets must be equal zero 
identically. Now we want to check this fact. The first Poisson bracket is
\begin{eqnarray}
 &[&N, \Pi^{mn} ] = [ \frac{1}{\sqrt{- g^{00}}}, p^{mn} ] = - \frac{1}{\sqrt{(- g^{00})^{3}}} [ g^{00}, p^{mn} ] \nonumber \\
 &=& 
 \frac{1}{\sqrt{(- g^{00})^{3}}} \frac12 ( g^{0 m} g^{0 n} +  g^{0 n} g^{0 m} ) = \frac{1}{\sqrt{(- g^{00})^{3}}} g^{0 m} g^{0 n} \ne 0
 \; , \; \label{N} 
\end{eqnarray}
while for the second bracket one finds
\begin{eqnarray}
 &[&N^{k}, \Pi^{mn} ] = [ -\frac{g^{0k}}{g^{00}}, p^{mn} ] = \frac{1}{2 g^{00}} ( g^{0 m} g^{0 n} + g^{0 n} g^{0 m} ) - 
 \frac{1}{( g^{00} )^{2}} g^{0 k} g^{0 m} g^{0 n} \nonumber \\
 &=&  \frac{1}{2 ( g^{00} )^{2}} ( g^{0 0} g^{0 m} g^{k n} + g^{0 0} g^{0 n} g^{k m}
 - 2 g^{0 k} g^{0 m} g^{0 n} ) \ne 0 \; , \; \label{N-k} 
\end{eqnarray}
where $k = 1, 2, 3$. So, I am sorry to say, but none of these four Poisson brackets equal zero identically. Therefore, these dynamical variables are 
not canonical and theory which uses these variables is not a Hamiltonian theory. Furthermore, it cannot be transformed into such a theory by any 
correct procedure and/or by applying any canonical transformation. Now, we can only guess that P.A.M. Dirac calculated these four Poisson brackets 
in the end of 1950's. Very likely, he was trying to say something to that "enthusiastic group of young fellows" (he worked in Frorida at that time), 
but those fellows simply ignored all his comments and doubts about their new and `far-advanced' Hamiltonian formulation of the metric GR. Finally, 
these yong authors created the new `super-advanced' geometro-dynamics, which later was called (and considered) by Hawking \cite{Hawk} as a theory 
which "contradicts to the whole spirit of General Relativity". However, such a contradiction is only a small problem for geometrodynamics, which 
proved to be incorrect and incomplete in its applications to the real problems of metric gravity (more details can be found in \cite{KK2011}).


\begin{thebibliography}{99}
\bibitem{QL} The coefficients in such `quasi-linear' combinations are field-dependent, i.e., they are neither numbers, nor `group' constants.
\bibitem{constr} It can be shown that all primary $\phi^{0\lambda}$ and secondary constraints $\chi^{0\sigma}$ which arise during this 
Hamiltonian procedure are the first class constraints \cite{Dir50}.
\end{thebibliography}
\end{document}